\documentclass[twocolumn,showpacs,preprintnumbers,amsmath,amssymb,aps]{revtex4}

\usepackage{graphicx}
\usepackage{dcolumn}
\usepackage{bm}
\usepackage{epstopdf}

\begin{document}

\title{Diffusion of self-propelled Janus tracer in polymeric environment}


\author{Nairhita Samanta, Rohit Goswami}

\author{Rajarshi Chakrabarti}%
 \email{rajarshi@chem.iitb.ac.in}
\affiliation{%
Department of Chemistry, Indian Institute of Technology Bombay, Powai, Mumbai 400076, India
}

\date{\today}

\begin{abstract}

Artificially synthesized Janus particles have tremendous prospective as in-vivo drug-delivery agents due to the possibility of self-propulsion by external stimuli. Here we report the first ever computational study of translational and rotational motion of self-propelled Janus tracers in a heterogeneous polymeric environment.  The presence of polymers makes the translational mean square displacement (MSD) of the Janus tracer to grow very slowly as compared to that of a free Janus tracer, but surprisingly the mean square angular displacement (MSAD) is significantly increased as observed in a recent experiment. Moreover, with the increasing propulsion velocity, MSAD grows even faster. However, when the repulsive polymers are replaced with polymers with sticky zones, MSD and MSAD both show sharp decline.

\end{abstract}


\maketitle

The presence of innumerable active matter systems in nature, inspired researchers to come up with innovative designs of smart artificial devices that can be made to self-propel \cite{sennanoscale2013, stadler}. Major number of these man-made devices are Janus particles. Janus particles are either micro or nano-particles, synthesized in such a way that the two halves of its surface possess two distinct physical or chemical properties and can interact with the environment anisotropically. The self-propulsion of the Janus particles is achieved either by coating half of its surface with a catalytic patches like Platinum which in turn help in catalyzing chemical reaction in its immediate environment. The chemical reaction causes a concentration gradient of the reactant and the product in the system, thus helping the particle to propel \cite{wu, stadler}. Another technique involves Laser illumination of the Janus particle that is placed in a critical Binary mixture. In this case, one half of the Janus particle is coated with a heavier metal and when shined with Laser, the particle propels on its heavier side \cite{Roijprl, bechingerprl}. There is huge potential for these artificial swimmers in the field of medicine. The most practical application involves targeted drug delivery and bio-sensing \cite{sennanoscale2013,wu, tom, katuri}. While there have been a significant progress in theoretical and experimental research on active systems, most of these studies focus on collective motion of active matter in homogeneous medium. However, the real systems where these artificial devices would be applied for drug delivery is crowded \cite{metzlerpccp,chakrabartipre2013,sokolov2012} and heterogeneous like biological cells \cite{metzlerphysto,bechingerreview}.

In this work we investigate the dynamics of an active Janus tracer in crowded environment. While there exist many theoretical and computational studies on the Janus tracer or Janus rod \cite{pulaksoftmat,sebastian,lowenjpcondmat,snigdhapre15,sen,winkler}, no attempt has been made so far to address the effect of viscoelastic medium on self-propelled Janus particle, excluding a very recent experiment \cite{bechingerprl}. Hence, to inspect the effect of complex viscoelastic environment we consider a Janus particle in a polymeric solution and employ computer simulations to analyze the dynamics.

\begin{figure}[h]
\centering
 \begin{tabular}{@{}cccc@{}}
 \includegraphics[width=.2\textwidth]{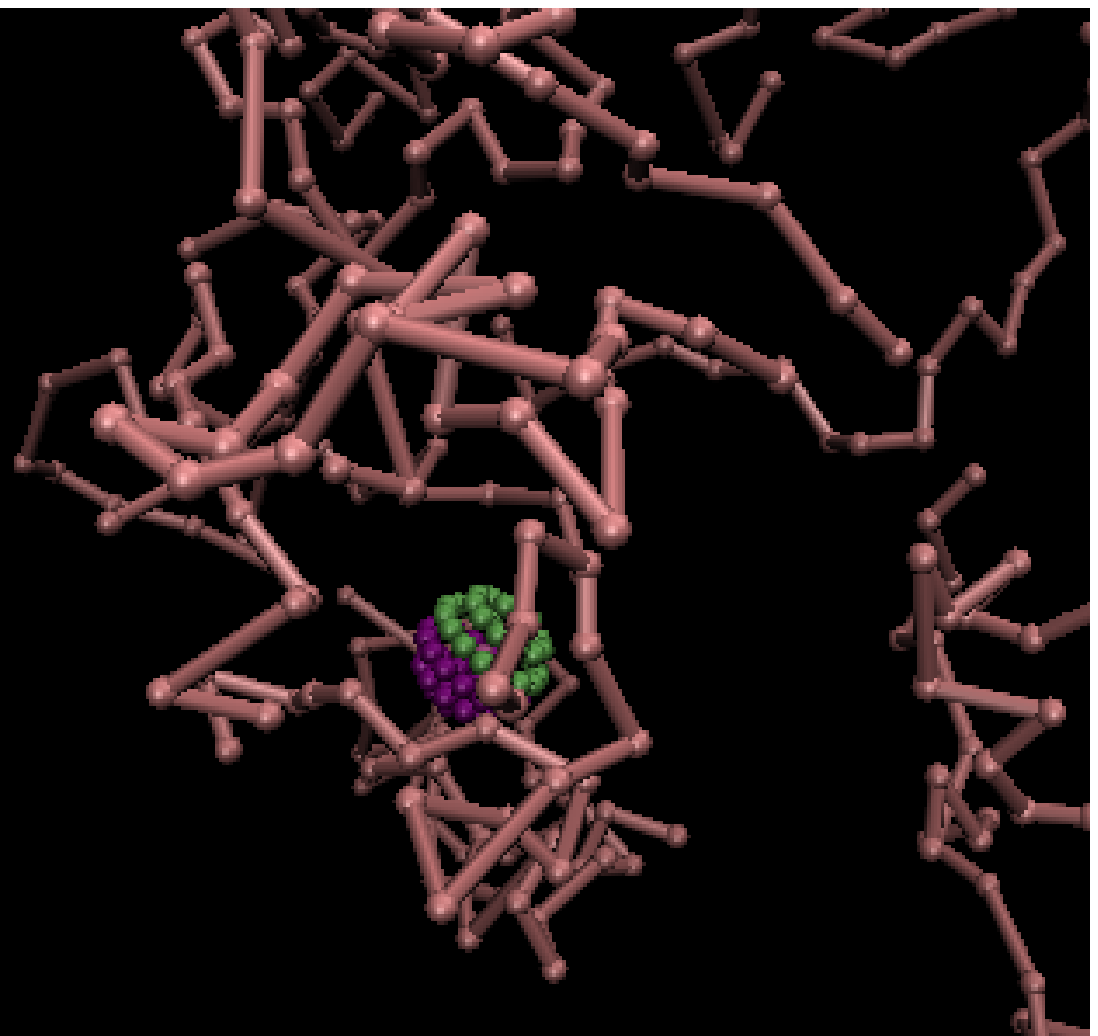}(a)
 \includegraphics[width=.2\textwidth]{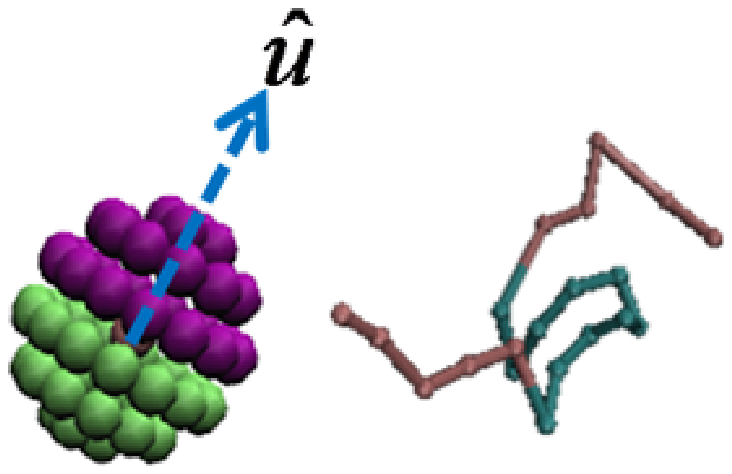}(b)
 \end{tabular}
 \caption{\label{fig:snapshot} (a) A typical snapshot of Janus particle in the presence of (repulsive, case I) polymers. (b) Schematic diagrams of a Janus tracer and a sticky polymer (case II). $\hat{u}$ is the orientational unit vector of the Janus particle and the green half of the Janus particle have attractive interaction with the cyan part of the polymers. }
 \end{figure}

The simulations are done in ESPResSo, a freely available software \cite{holmespresso}. We use Lennard-Jones parameters as the unit where $\epsilon_0$ and $\sigma_0$ are the unit of energy and length respectively. The simulation box is cubic with a size of $10\sigma_0$. The polymers in the system as shown in Fig. (\ref{fig:snapshot}a), has twenty monomers each of the size $0.5\sigma_0$. The neighboring monomers of the polymer are connected by FENE potential $V_{FENE}=-\frac{k_f r^2_{max}}{2}log\left[1-\left(\frac{r}{r_{max}}\right)^2\right]$. The values of the parameters considered here are $k_f=7$ and $r_{max}=2$.

Our Janus particle is spherical in structure and it is created by covering the surface of one sphere with another set of small spheres. As shown in Fig. (\ref{fig:snapshot}b) the spheres on the surface consist of two different types and are equal in number. This results in two distinct halves of the Janus particle consisting of thirty particles each. These two halves can interact with the surroundings differently. The diameters of the central sphere and the spheres on the surface are same, i.e. $0.5\sigma_0$. The distance of the surface particle from the centre is $0.5\sigma_0$, which makes the radius of the Janus particle to be $0.75\sigma_0$.

\noindent Here we consider two cases. Firstly one with purely repulsive polymers and the other is the one with polymers having sticky zones.

\noindent \textbf{Case I:} We consider the interaction of the Janus particle with the polymers in the system completely repulsive with soft-core Weeks-Chandler-Anderson (WCA) potential, $V_{WCA}(r)=4\epsilon\left[\left(\frac{\sigma}{r}\right)^{12}-\left(\frac{\sigma}{r}\right)^{6}\right]+\epsilon, \mbox{if } r<(2)^{1/6}\sigma$, otherwise $V_{WCA}(r)=0$ and $\epsilon=1\epsilon_0$.


\noindent \textbf{Case II:} We introduce trapping zone to each polymer \cite{chakrabarti}. These trapping zones can bind the Janus particle due to attractive interaction between one half of the Janus particle and them. The trapping zone of each polymers consists of ten monomers starting from $6^{th}$ to $15^{th}$. A representation of the this sticky polymer can be seen in Fig. (\ref{fig:snapshot}b). The interaction between this sticky part of the polymer and one half of the Janus tracer is modelled by Lennard-Jones (LJ) potential. $V_{LJ}(r)=4\epsilon\left[\left(\frac{\sigma}{r}\right)^{12}-\left(\frac{\sigma}{r}\right)^{6}\right], \mbox{if } r<r_{cut}$ and $V_{LJ}(r)=0$ when, $r>r_{cut}$.
The values of the parameter considered here, $\epsilon=2\epsilon_0$, $r_{cut}=3$. The rest of monomers in the polymers remains repulsive to both halves of the Janus particle. As before, the repulsive interaction is modelled by WCA potential. To avoid collapse of the polymers on each other or on itself we employ soft core WCA interaction between all the monomers. The general governing equation of motion for all the passive particles is the following Langevin equation, $m\frac{d^2 r_i(t)}{dt^2}=-\xi\frac{dr_i}{dt}-\bigtriangledown\sum_{j} V(r_i-r_{j})+f_i(t)$. Where, $m$ is the mass of the particle and $\xi$ is the friction coefficient. In all the simulations $\xi$ is considered to be $1$. $V$ is the pair potential can either be LJ or WCA.
$r(t)$ is the position vector and $f(t)$ is the thermal noise with zero first moment:
 $\left<f(t)\right>=0,
   \left<f_{\alpha}(t^{\prime})f_{\beta}(t^{\prime\prime})\right>\sim \xi k_B T \delta_{\alpha\beta}\delta(t^{\prime}-t^{\prime\prime})$.
   All the simulations are performed using Langevin thermostat in NVT ensemble where $T$ is the temperature of the bath, $k_B$ is the Boltzmann constant,  and thermal energy $k_BT=1$. The dynamics of the Janus particle is modified by introducing a driving force $f_p$ in the Langevin equation, that eventually gives rise to a propulsion velocity $v=\frac{f_p}{\xi}$. This additional force accounts for the self-propulsion. Thus for the self-propelled Janus particle the equation of motion is same as the above with an added term $f_p\hat{u}$. Where, $\hat{u}$, orientational unit vector of the Janus particle as shown in Fig. (\ref{fig:snapshot}b). At the beginning of the simulation i.e. at $t=0$, $\hat{u}$ is directed towards the $z$-axis. $\hat{u}$  basically acts here as a body-frame axis which can undergo translation and rotation along with the Janus tracer due to the thermal fluctuation. The propulsion velocity always acts along the direction of $\hat{u}$ \cite{joostjcp}. Since the orientational unit vector can rotate, there exists another equation of motion for the Quaternion of the Janus particle \cite{mountain}. The simulations are run for $2\times10^6$ steps where the time steps $dt$ is considered to be $0.001$.

\onecolumngrid
\begin{center}
\begin{figure}[h]
 \begin{tabular}{@{}cccc@{}}
 \includegraphics[width=0.35\textwidth]{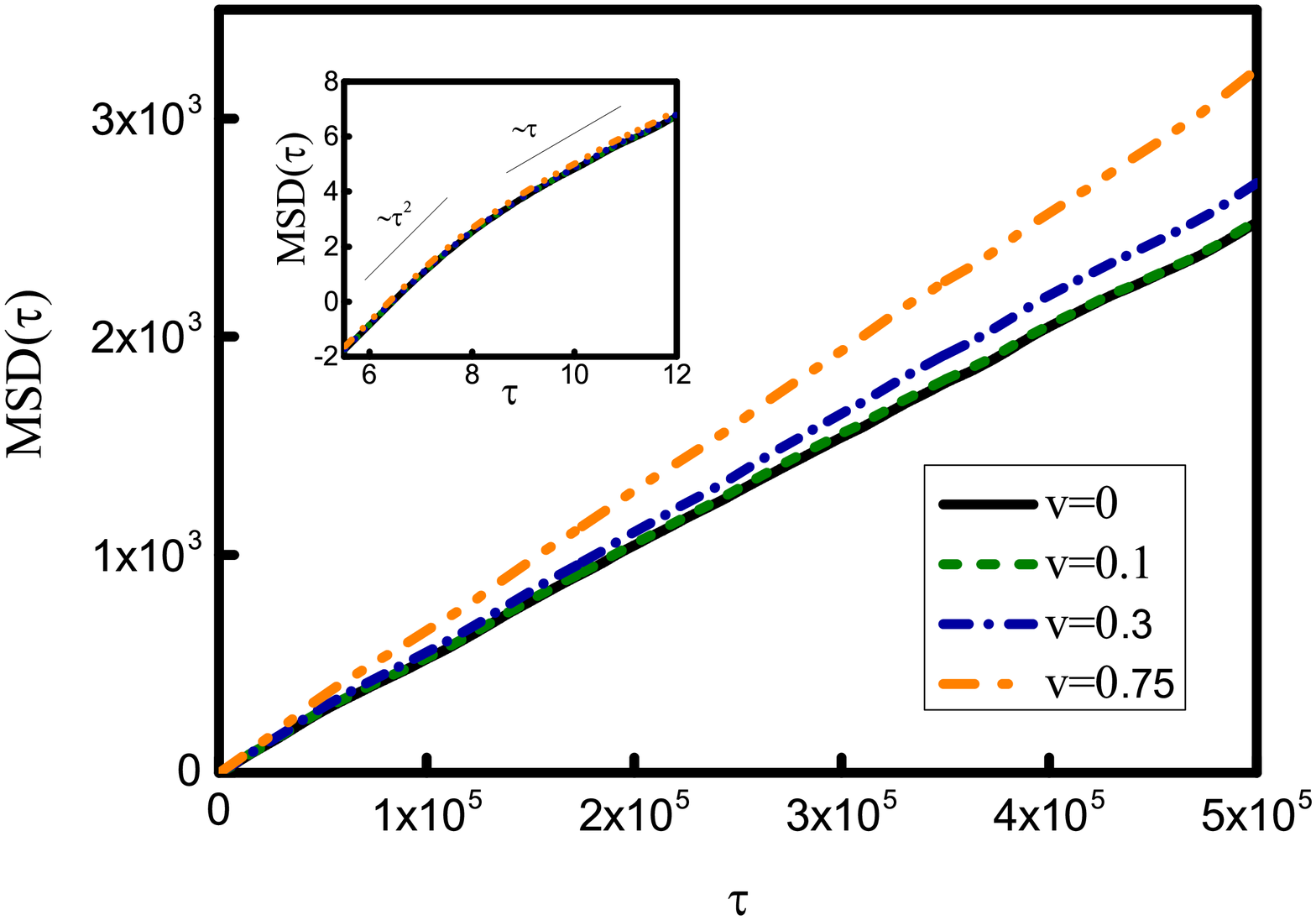}(a)
 \includegraphics[width=0.35\textwidth]{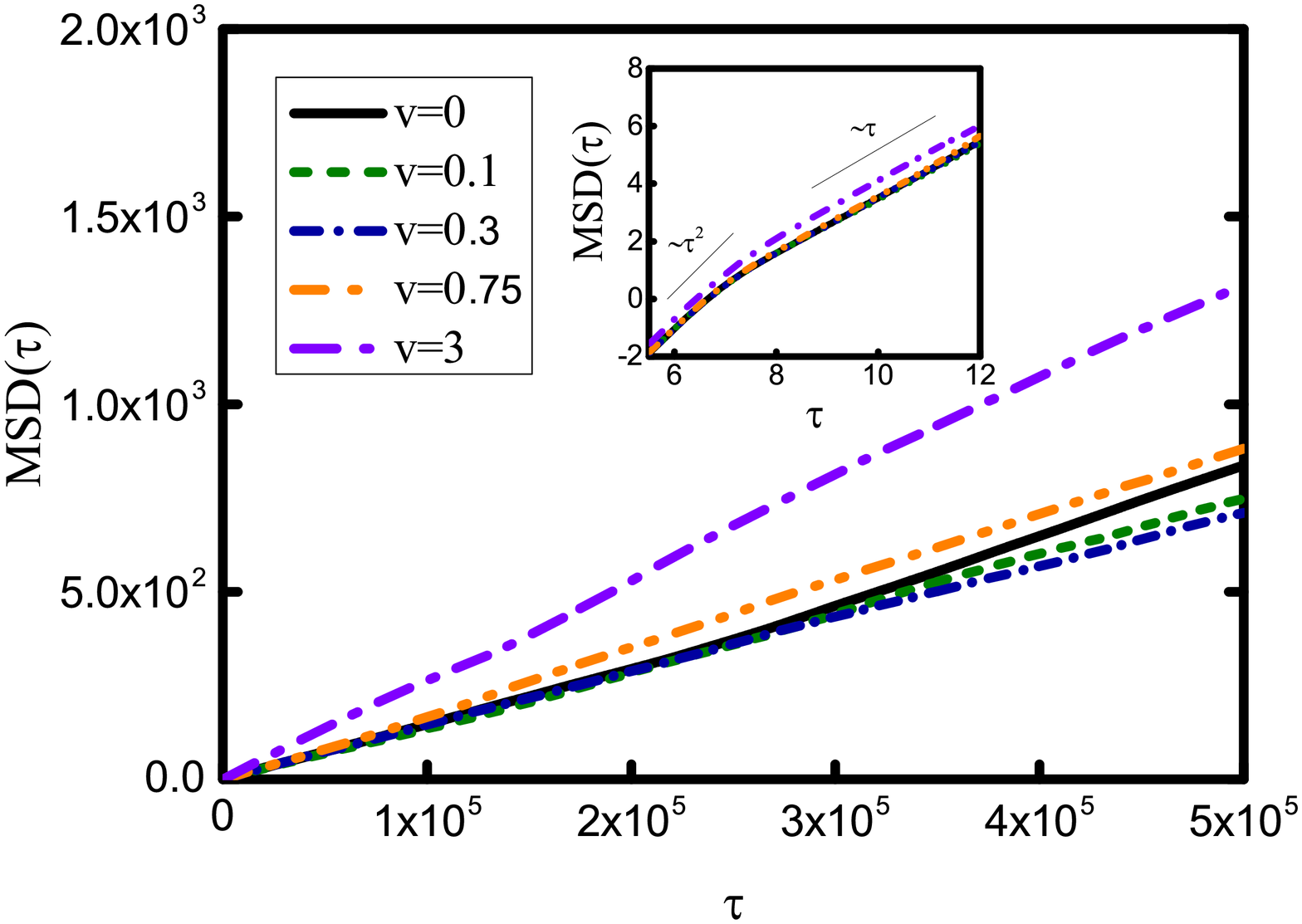}(b)
 \end{tabular}
 \caption{\label{fig:msdc} MSD plots for the (a) free Janus tracer (b) Janus tracer in polymeric environment (repulsive, case I) as a function of the elapsed time at different propulsion velocity $v$. Insets show the log-log (natural) plot for the same. }
 \end{figure}
 \end{center}
\twocolumngrid

Fig. (\ref{fig:msdc}a) shows the time and ensemble averaged mean square displacement (MSD), $MSD=\overline{\delta^{2}(\tau)}=\overline{[r(t+\tau)-r(t)]^2}$ of a free Janus particle in the absence of any crowding. The simulation are carried out for different propulsion velocity and compared with that of a passive Janus tracer ($v=0$). As can be seen from the log-log plot in the inset at short time the particle passes through a ballistic regime and then crosses over to normal diffusion. As expected the MSD of an active Janus tracer grows faster in comparison to that of a passive one and with increasing propulsion velocity the MSD shows a steady increase. Next, to study the effect of crowding we introduce polymers in the system and carry out simulations at different propulsion velocity of the Janus particle. The volume fraction of the system is $5\%$. The volume fractions is defined as the ratio of the volume occupied by the total number of particles in the system to the box volume. Here, the interaction between the polymer and the tracer is purely repulsive. Fig. (\ref{fig:msdc}b) shows the MSD of the Janus tracer in the presence of repulsive polymers and we see a sharp decrease in the order of magnitude for MSD when compared with that of the free Janus. Here also the MSD grows ballistically at first and then become diffusive. However, in the presence of crowding, MSD does not grow that fast for small propulsion velocity as can be seen for the free Janus particle. Even for $v=0.75$, MSD is quite similar as that of the passive Janus particle and we do not see a proper trend of the increasing MSD as propulsion velocity becomes higher. However, if one increases the propulsion velocity even further, the increment of the MSD becomes visible. For reference, we have shown the MSD for the Janus particle with propulsion velocity of $v=3$. At this high propulsion velocity MSD grows much faster as compared to the purely repulsive case, $v=0$.

In addition to the translational dynamics we assess the rotational motion of the Janus particle and calculate Mean square angular displacement (MSAD) for that. To evaluate the MSAD we closely follow the prescription given by Mazza \textit{et. al.} \cite{stanleyprl}. First an unit vector $\hat{p}$ is defined and that is chosen to be same as the orientational unit vector $\hat{u}$. The Rotation matrix $\textbf{R}$ for the Janus tracer is then calculated to find $\hat{p}(t)$ from $\hat{p}(0)$ as done by Hunter \textit{et. al.} earlier \cite{weeksopticex}. $\hat{p}(0)$ is directed along $z$-axis at $t=0$ and $\hat{p}(t)=\prod_k{\textbf{R}^k}\hat{p}(0)$. The rotational displacement $\vec{\phi}(t)$ is obtained from the rotational velocity $\vec{\omega}(t)$, i.e. $\vec{\phi}(t)=\int_0^t{\vec{\omega}(t^{\prime})}dt^{\prime}$. Where, the direction of $\vec{\omega}(t)$ is obtained from $p(t^{\prime})\times p(t^{\prime}+dt^{\prime})$ and the magnitude is $|\vec{\omega}(t^{\prime})|=cos^{-1}[\hat{p}(t^{\prime}).\hat{p}(t^{\prime}+dt^{\prime})]$. Finally we calculate the time and ensemble averaged MSAD from $\vec{\phi}(t)$ in a similar way we calculate translational MSD, $MSAD=\overline{[\vec{\phi}(t+\tau)-\vec{\phi}(t)]^2}$.

\begin{figure}
\centering
 \includegraphics[width=.35\textwidth]{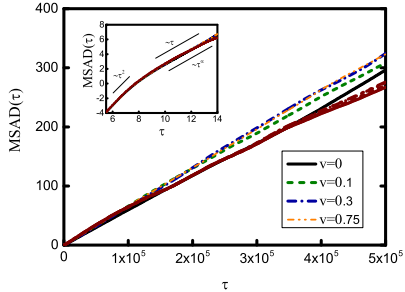}
 \caption{\label{fig:msadc} MSAD plots for the Janus tracer in polymeric environment (repulsive, case I) as a function of the elapsed time at different propulsion velocity $v$. MSAD plots for the free Janus tracer is shown in wine color for all the propulsion velocity with similar line style as that of Janus in polymeric environment. Inset shows the log-log (natural) plot for the same.}
 \end{figure}

\noindent In Fig. (\ref{fig:msadc}) we show the MSAD for the free Janus as well as the MSAD of the same Janus particle in the visco-elastic polymeric medium. For MSAD, the ballistic regime can also be clearly distinguished at short time. The free Janus particle shows very weak subdiffusive (MSD$\sim\tau^{\alpha}$, $\alpha \sim0.9$) behavior at long time. We observe in the absence of polymers, MSAD does not vary significantly and it practically shows no dependence on $v$. Even from the plots it is very difficult to distinguish the MSAD for different $v$. Surprisingly, in the presence of polymers the particle although begins with the ballistic motion, it turns to be diffusive in long time. One would also expect the MSAD for the Janus tracer to grow slowly in the visco-elastic medium if compared with the MSAD of the free Janus particle. On the contrary in the presence of polymers, the MSAD for the Janus tracer grows faster than that of the free one. Moreover, with increasing propulsion velocity the MSAD shows a steady increase. Although intuitively it is expected that with higher propulsion velocity the MSAD would grow slowly as the particle would propel efficiently, i.e. it will translate more and rotate less. This would happen as the particle becomes self-propelled it would be less influenced by the thermal noise from the environment. Hence our observation is undoubtedly very startling. As a matter of fact, recently similar trend has been observed by Gomez-Solano \textit{et. al} in an experiment, where they noticed an enhancement of the MSAD for a Janus particle in viscoelastic medium in contrast to Newtonian fluid \cite{bechingerprl}. Such increment of MSAD is the consequence of the collisions between the Janus tracer and the polymers and these collisions eventually help the tracer to rotate. This effect is completely absent in simple fluid. We expect in the presence of polymers as well if the propulsion velocity becomes very high, the MSAD will show a decline as the particle would translate more in that circumstances and collision from the polymers would have less impact on it.

\onecolumngrid
\begin{center}
\begin{figure}[h]
\centering
 \begin{tabular}{@{}cccc@{}}
 \includegraphics[width=0.35\textwidth]{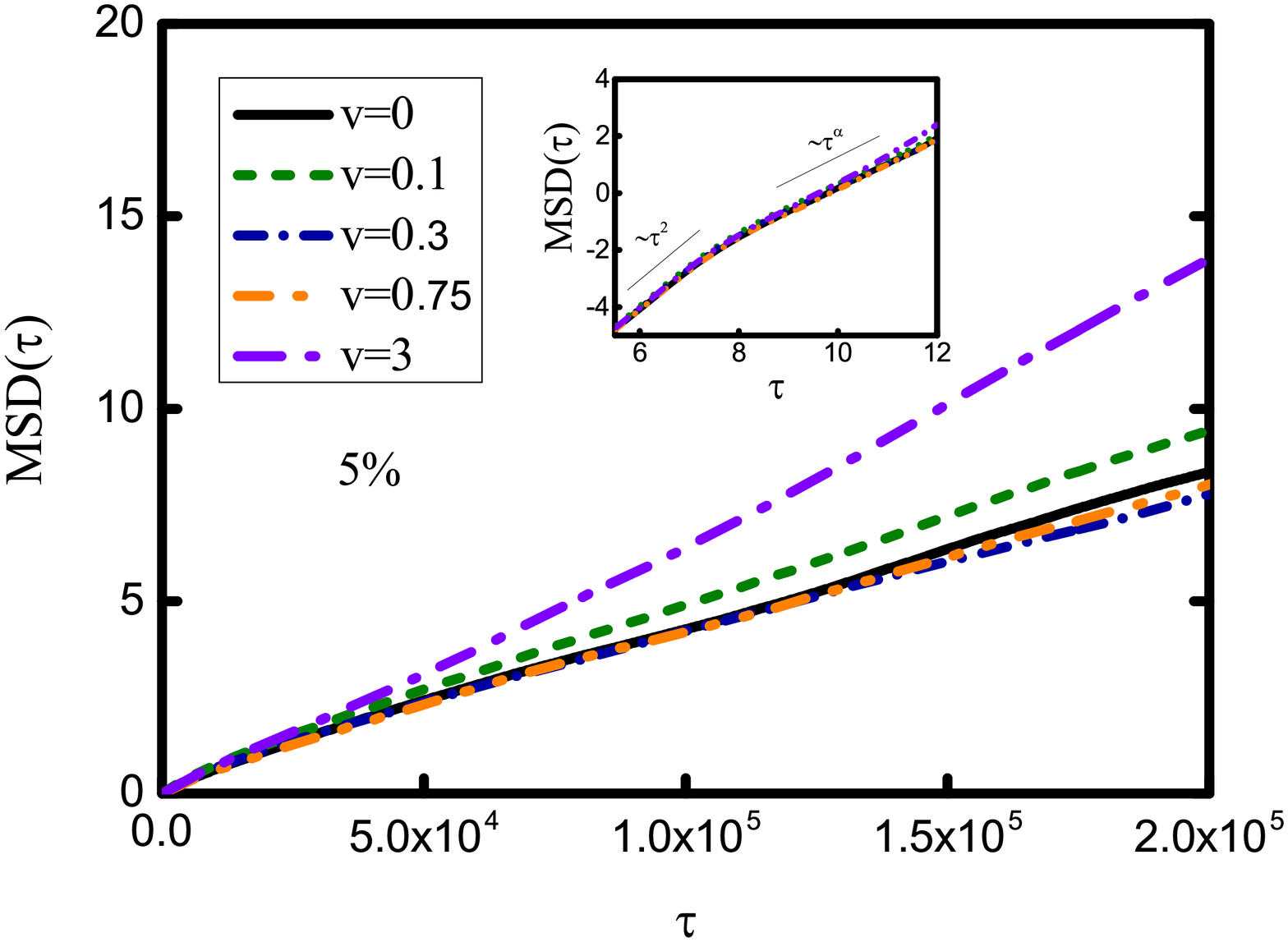}(a)
 \includegraphics[width=0.35\textwidth]{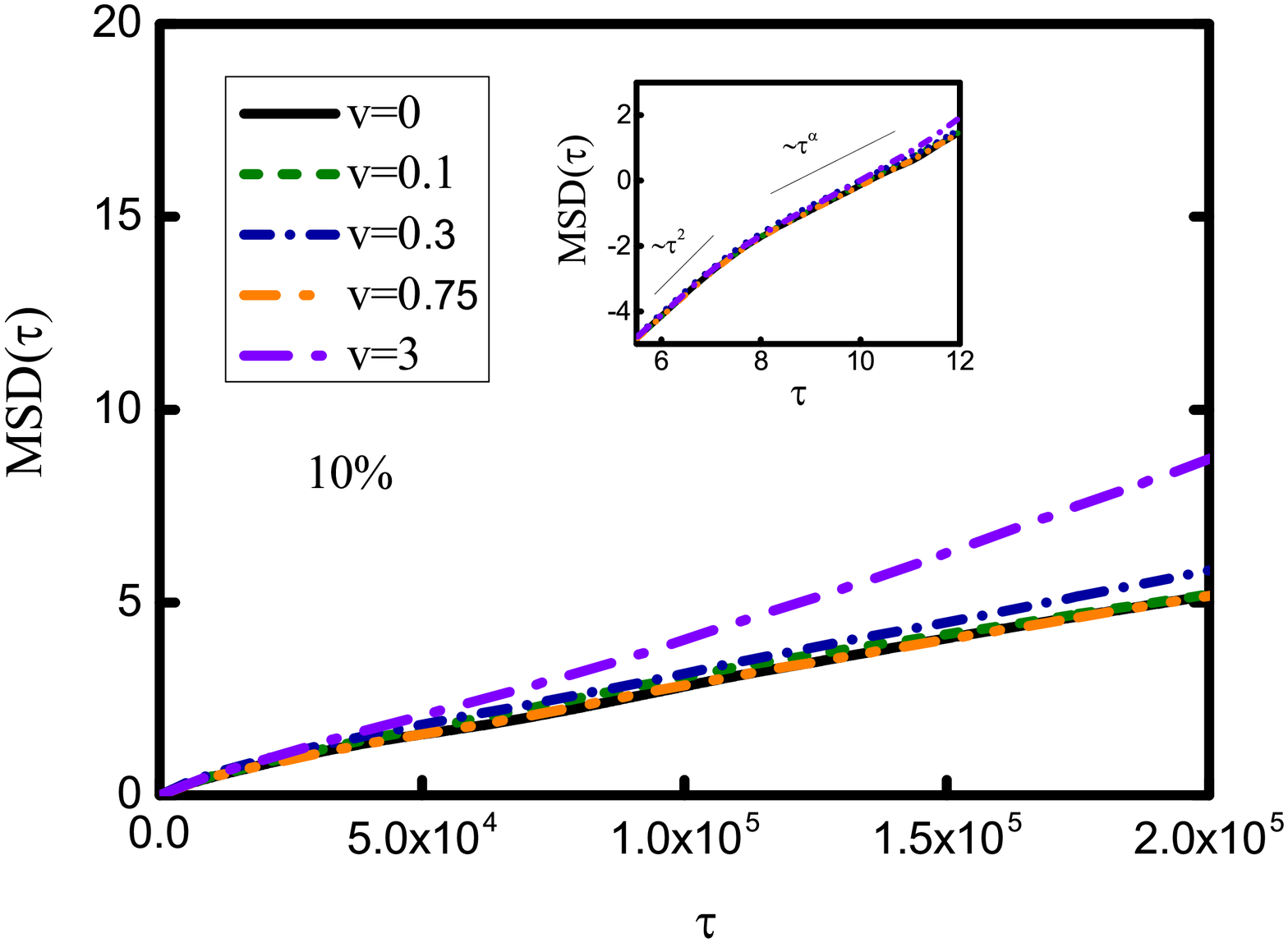}(b)
 \end{tabular}
 \caption{\label{fig:msds} MSD plots for the Janus tracer as a function of the elapsed time at different propulsion velocity $v$ in the presence of sticky polymers (case II) at (a) $5\%$ (b) $10\%$ volume fractions respectively . Insets show the log-log (natural) plot for the same.}
 \end{figure}
 \end{center}
 \twocolumngrid

To probe the effect of viscoelastic medium further, we incorporate sticky polymers in the system instead of the simple repulsive ones (case II). We calculate the translation MSD and the rotational MSAD. Fig. (\ref{fig:msds}) shows the MSD for the Janus particle in the sticky polymeric environment. The simulations are executed in two different volume fractions to assess the effect of crowding along with the stickyness of the polymers. Fig. (\ref{fig:msds}a) and Fig. (\ref{fig:msds}b) are at volume fractions $5\%$ and $10\%$ (semi-dilute regime) respectively. The tracer passes through a ballistic region at short time and in long time it exhibits subdiffusion in both the cases. In addition to that MSD for the tracer in the presence of sticky polymers grows very slowly in sharp contrast to the MSD in the presence of repulsive polymers. There is two fold decrease in the order of magnitude. This happens as the Janus particle gets surrounded by the polymers that eventually prevent the tracer movement and even at higher propulsion velocity the tracer cannot overcome the effect of the sticky polymeric environment. With the increase of the crowding the MSD is expected to decline and that exactly happens when the volume fraction is increased. Although the change is very small, by a factor of $\sim1.5$ times when the extent of crowding is increased by a factor of $2$. In both the cases, for smaller propulsion velocity the MSD of self-propelled Janus tracer ($v\neq0$) grows very similarly as that of the passive one ($v=0$). However for even higher propulsion velocity ($v=3$), distinct enhancement of MSD is noticed in both the cases.

\onecolumngrid
\begin{center}
 \begin{figure}[h]
\centering
 \begin{tabular}{@{}cccc@{}}
 \includegraphics[width=0.35\textwidth]{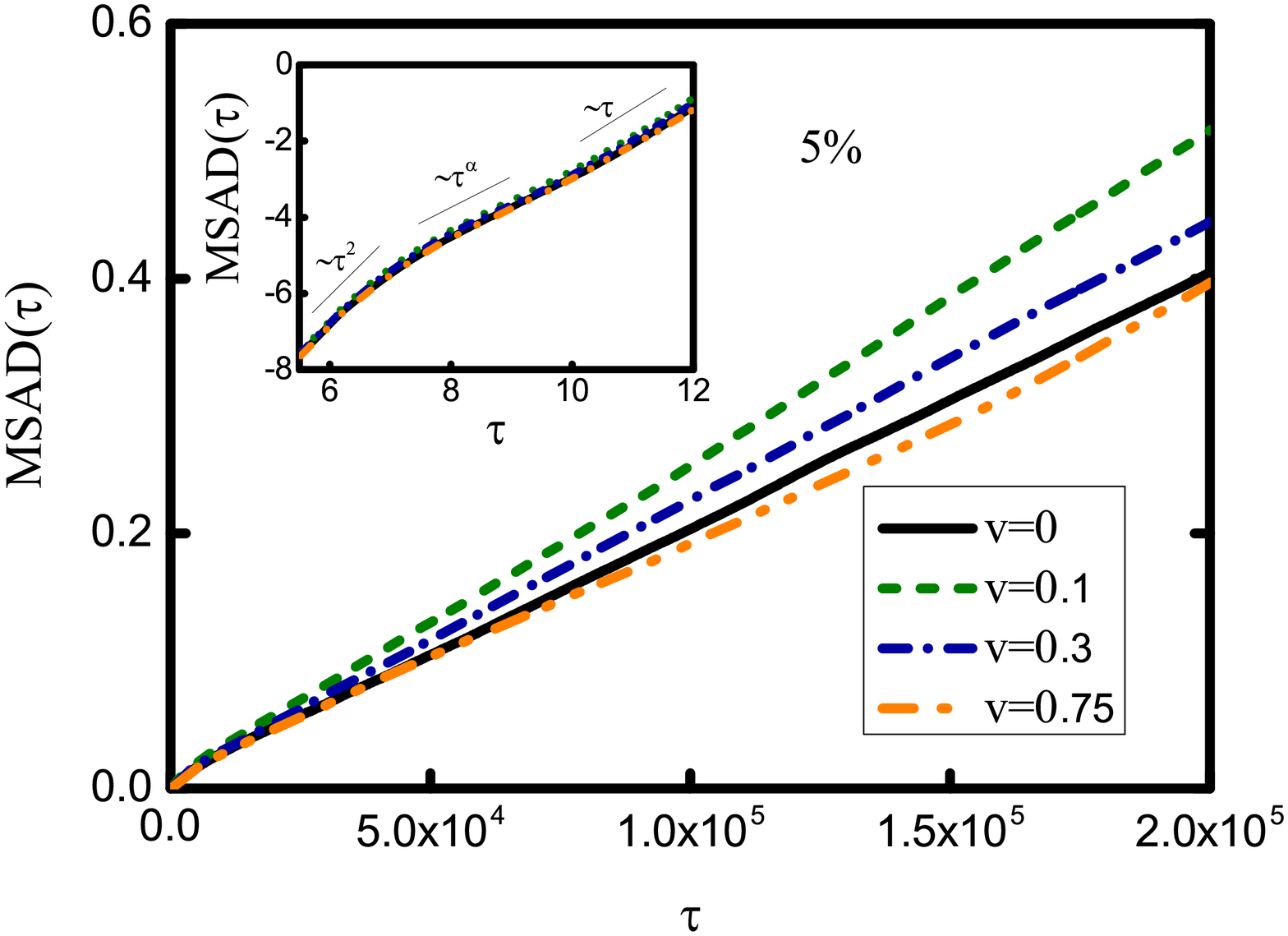}(a)
 \includegraphics[width=0.35\textwidth]{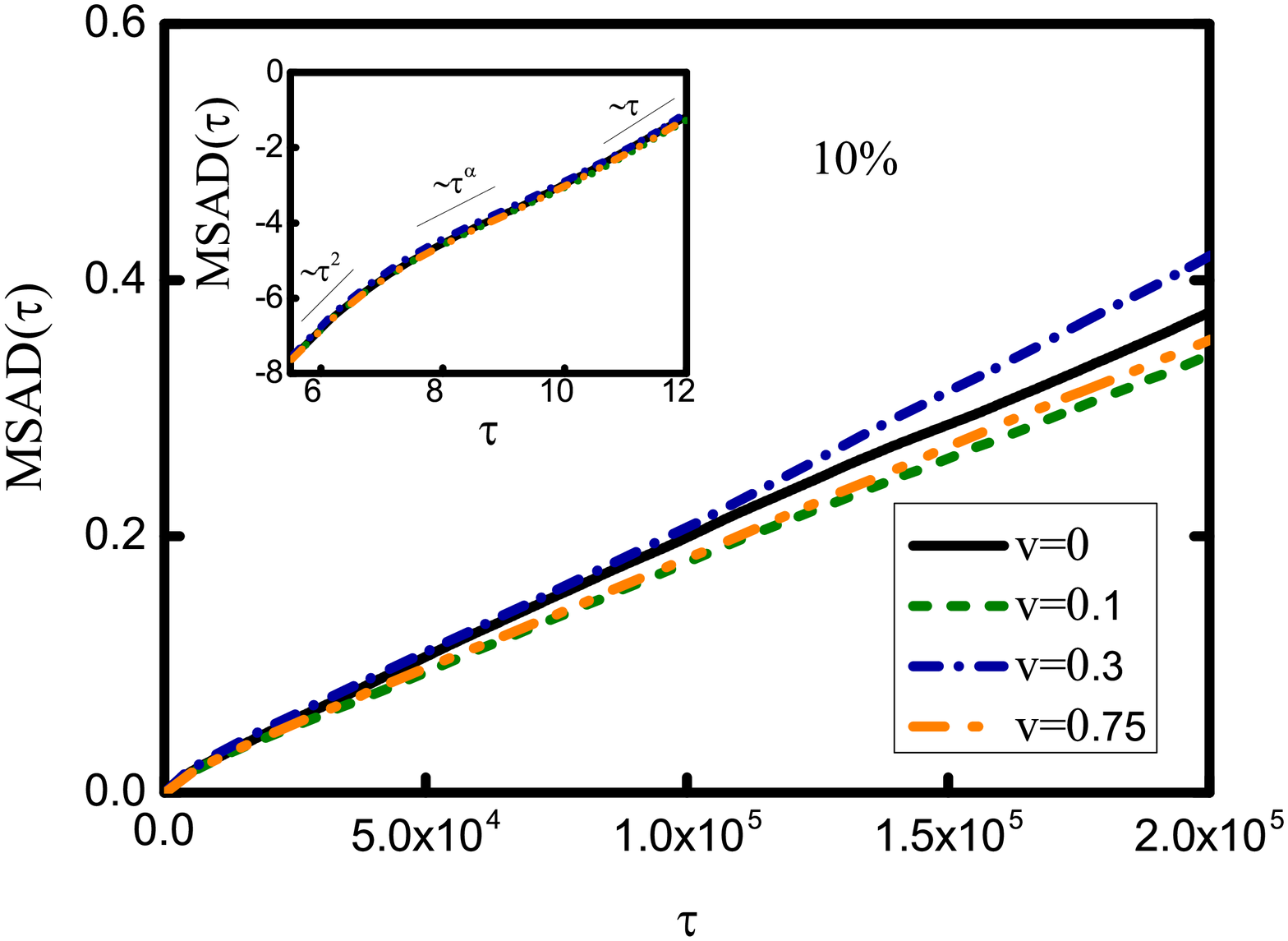}(b)
 \end{tabular}
 \caption{\label{fig:msads} MSAD plots for the Janus tracer as a function of the elapsed time at different propulsion velocity $v$ in the presence of sticky polymers (case II) at (a) $5\%$ (b) $10\%$ volume fractions respectively . Insets show the log-log (natural) plot for the same.}
 \end{figure}
 \end{center}
\twocolumngrid
Fig. (\ref{fig:msads}a) and Fig. (\ref{fig:msads}b ) show the MSAD of the same Janus tracer in the presence of sticky polymers at different degrees of crowding. From the log-log plots (inset) of the MSAD against elapsed time it can be seen that in both the cases the Janus tracer starts off with the ballistic regime and becomes diffusive at long time after passing through an intermediate subdiffusive region. Although MSAD shows a huge decline on inclusion of the sticky zones in the polymers, it does not get influenced much by the crowding. As even at the two fold increase of the volume fraction, there is no substantial change in the the MSAD when we go from $5\%$ to $10\%$ volume fraction of the polymers. However here we see a very interesting trend of the MSAD when the propulsion velocity is varied. In case of tracer at $5\%$ volume fraction, the MSAD grows slightly faster when particle becomes self-propelled. However, when the propulsion velocity is increased even further the growth of the MSAD diminishes. Therefore, at smaller propulsion velocity although the collisions from the polymers help to rotate the Janus particle, at higher propulsion velocity the effect of the propulsion velocity overcomes the effect of the polymer and MSAD decreases. For the higher volume fraction we see similar behavior for MSAD. At $v=0.1$, the MSAD is comparable to that of the passive Janus particle. With increasing propulsion velocity MSAD grows little faster (for $v=0.3$) and on further increment of $v$ it becomes slower.

In a nutshell, we have studied the effect of viscoelastic medium on the translational and rotational dynamics of a self-propelled Janus particle. The results show that although there is an offset of translational MSD of a self-propelled tracer in the presence of polymers, the MSAD surprisingly grows faster in the presence of repulsive polymers. Our result for the rotational motion is in the same spirit of the recent experimental observation \cite{bechingerprl}. However in the presence of sticky polymers both translational and rotational motion of the tracer get affected. Although the MSAD grows a little faster when the propulsion velocity of the tracer is very small, at higher propulsion velocity MSAD shows a steady decline. Our observations are important and have implications in understanding the dynamics of self-propelled drug-delivery vehicle in crowded environment such as biological cells.

We acknowledge SERB (Project No. SB/SI/PC-55/2013) for funding.

\end{document}